\documentclass{aa}
\usepackage[dvips]{graphicx}
\usepackage{mathrsfs}
\usepackage{amsmath}
\usepackage{amssymb}
\usepackage[mathscr]{eucal}
\usepackage{latexsym}
\usepackage{amsbsy}    
\usepackage{float}
\usepackage[latin1]{inputenc}

\newcommand{\bd}{\mathbf}

\def\Real{{\rm I\mathchoice{\kern-0.70mm}{\kern-0.70mm}{\kern-0.65mm}%

  {\kern-0.50mm}R}}

\font \bolditalics = cmmib10

\def\bx#1{\leavevmode\thinspace\hbox{\vrule\vtop{\vbox{\hrule\kern1pt

        \hbox{\vphantom{\tt/}\thinspace{\bf#1}\thinspace}}

      \kern1pt\hrule}\vrule}\thinspace}

\def \vc #1{{\textfont1=\bolditalics \hbox{$\bf#1$}}}

\def\thetag{{\vc \theta}}

\def\Cg{{\bf C}}

\def\be{\begin{equation}}
\def\ee{\end{equation}}
\def\ba{\begin{eqnarray}}
\def\ea{\end{eqnarray}}

\def\ltsima{$\; \buildrel < \over \sim \;$}
\def\lsim{\lower.5ex\hbox{\ltsima}}
\def\gtsima{$\; \buildrel > \over \sim \;$}
\def\gsim{\lower.5ex\hbox{\gtsima}}

\begin{document}







\title{Cosmic Shear Analysis with CFHTLS Deep data\thanks{Based on
observations obtained with {\sc MegaPrime/Megacam}, a joint project of
CFHT and CEA/DAPNIA, at the Canada-France-Hawaii Telescope (CFHT)
which is operated by the National Research Council (NRC) of Canada,
the Institut National des Science de l'Univers of the Centre National
de la Recherche Scientifique (CNRS) of France, and the University of
Hawaii. This work is based in part on data products produced at {\sc
Terapix} and the Canadian Astronomy Data Centre as part of the
Canada-France-Hawaii Telescope Legacy Survey, a collaborative project
of NRC and CNRS.}}

\author{E. Semboloni$^{1}$, Y. Mellier$^{1,2}$, L. van Waerbeke$^{3}$,
H. Hoekstra$^{4}$, I. Tereno$^{1,6}$, K. Benabed$^{1,2}$,
S.D.J. Gwyn$^{4}$, L. Fu$^{1}$, M.J. Hudson$^{7}$, R. Maoli$^{1,5}$,
L.C. Parker$^{7}$} \offprints{sembolon@iap.fr}

\institute{%
$^1$ Institut d'Astrophysique de Paris, UMR7095 CNRS, Universit\'e
Pierre \& Marie Curie, 98 bis boulevard Arago, 75014 Paris, France \\
$^2$ Observatoire de Paris. LERMA. 61, avenue de l'Observatoire.
75014 Paris, France.\\
$^3$ University of British Columbia, Department of Physics and
Astronomy, 6224 Agricultural Road, Vancouver, B.C. V6T 1Z1, Canada.\\
$^4$ Department of Physics and Astronomy. University of
Victoria. Victoria, B.C. V8P 5C2, Canada.\\
$^5$ Department of Physics. University La Sapienza. Pl. A. Moro 2,
00185, Roma, Italy. \\
$^6$ Departamento de Fisica. Universidade de Lisboa, 1749-016 Lisboa,
Portugal. \\
$^{7}$ Department of Physics. University of Waterloo, Waterloo ON N2L
3G1, Canada.
}

\markboth{Cosmic Shear Analysis in the CFHTLS Deep Fields}{}

\authorrunning{Semboloni et al.}
\date{ Received 5 November 2005, Accepted 31 January 2006}
\abstract{We present the first cosmic shear measurements obtained from
the T0001 release of the Canada-France-Hawaii Telescope Legacy
Survey. The data set covers three uncorrelated patches (D1, D3 and D4)
of one square degree each, observed in $u^*$, $g'$, $r'$, $i'$ and
$z'$ bands, to a depth of $i'=25.5$. The deep, multi-colour
observations in these fields allow for several data-quality
controls. The lensing signal is detected in both $r'$ and $i'$ bands
and shows similar amplitude and slope in both filters. $B$-modes are
found to be statistically zero at all scales.  Using multi-colour
information, we derived a photometric redshift for each galaxy and use
this to separate the background source sample into low-z and high-z
subsamples.  A stronger shear signal is detected from the high-z
subsample than from the low-z subsample, as expected from weak lensing
tomography.  While further work is needed to model the effects of
errors in the photometric redshifts, this result suggests that it will
be possible to obtain constraints on the growth of dark matter
fluctuations with lensing wide field surveys.  The combined
Deep and Wide surveys give $\sigma_8= 0.89 \pm 0.06$ assuming the
Peacock \& Dodds non-linear scheme (P\&D), and $\sigma_8=0.86 \pm
0.05$ for the halo model and $\Omega_m=0.3$. We assumed a Cold Dark
Matter model with flat geometry and have marginalized over the
systematics, the Hubble constant and redshift uncertainties. Using
data from the Deep survey, the $1\sigma$ upper bound for $w_0$, the
constant equation of state parameter is $w_0 < -0.8$.}
\maketitle
\section{Introduction}

Cosmological weak lensing, also called cosmic shear, can be used to
probe the dark matter distribution in the universe.  Weak lensing
observations complement other probes such as CMB anisotropies
(\cite{Sp03al}), type Ia supernovae (\cite{R04,R98,P99}), and redshift
surveys (\cite{lahavsuto04}). Weak lensing also has the advantage of
being free of any assumption regarding the light versus matter
distributions (\cite{ME99,BS01,VWM03,R03}).

It has only recently been shown that cosmic shear measurement is
technically feasible (\cite{BRE00,K00,VW00,WT00}). Unfortunately, the
deepest weak lensing survey has a sky coverage limited to less than
one deg$^2$ and the widest to $\sim$10 deg$^2$. Moreover, most surveys
were performed in one colour only, and even rough redshift information
was not available.  These limitations restricted the use of weak
lensing as a cosmological probe to a very small number of parameters.
Early weak lensing surveys were primarily focused on the measurements
of the normalization of the dark matter power spectrum, $\sigma_8$,
and the mass density parameter, $\Omega_m$.  The most recent cosmic
shear surveys reach a relative accuracy of about 10\% on
$\Omega_m\,\sigma_8^{0.5}$ (\cite{ME99,BS01,VWM03,R03}), but the
uncertainty on other parameters is still fairly large.

Second generation cosmic shear surveys are now under way and will
provide the community with multi-colour data of excellent image
quality, over a wide field of view. The Canada-France-Hawaii Telescope
Legacy Survey (CFHTLS)\footnote{{\tt http:
//www.cfht.hawaii.edu/Science/CFHLS/}}, using the recently built {\sc
MegaPrime/Megacam} wide field camera, belongs to this generation. The
{\it CFHTLS-Wide} survey (the core of the CFHTLS cosmic shear survey)
will provide a large sky coverage of 170 deg.$^2$, and the deep four
deg.$^2$ {\it CFHTLS-Deep} will provide shear information on smaller
scales and as a function of lookback time, out to higher redshift than
the {\it CFHTLS-Wide} .
 
Both surveys will ultimately consist of complete and homogeneous
panchromatic data in $u^*$,$g'$,$r'$,$i'$,$z'$. The data were taken
between June 1st 2003 and July 22, 2004 as part of a preliminary
survey to provide detailed quality assessments and propose technical
or operational improvements, when necessary.  The CFHTLS-Wide has the
same depth as {\sc Virmos}-Descart, but so far, the available data is
only in one colour and covers $\approx 20$ deg$^2$. A cosmic shear
analysis with the wide data is performed in \cite{HHal05}. The
CFHTLS-Deep has been observed in all five filters, therefore
photometric redshift are available, and will be used in this
work. Moreover, the CFHTLS D1 Deep field is located in the {\sc
Virmos}-VLT Deep Survey (VVDS) F02 field which has several thousand of
galaxy redshifts (\cite{Olf04}) and near infrared data (on a tiny
area). A combination of large and small scales from the Wide and the
Deep data will ultimately provide an excellent data set to probe the
nature of dark energy in the universe
(\cite{CoorHut99,BenBer01,LinJen03,BenVW04,JanvJain05}). In this work,
we describe the first CFHTLS cosmic shear studies based on Deep data,
and then combine the Wide and Deep data analysis to derive constraints
on $\Omega_m$ and $\sigma_8$.

The organization of the paper is as follows. In section 2 we introduce
the notation and define the statistics we use. The
data set is described in Section 3. In section 4 and 5 we present
results and residual systematics and we discuss them. Conclusions and
perspectives are outlined in Section 6.

\section{Theoretical background}

The theory of weak lensing has been previously been discussed in
detail in the literature, including the physical motivations of
various approximations (e.g. \cite{BS01}).  Following
\cite{H04,VWM02al,VWM05al} we follow the notation of \cite{SVWJK98}.

We introduce the power spectrum of the convergence $\kappa$ as :
\begin{eqnarray}
P_\kappa(k)&=&{9\over 4}\Omega_0^2\int_0^{\chi_H} {{\rm d}\chi \over
a^2(\chi)} P_{3D}\left({k\over f_K(\chi)};
\chi\right)\times\nonumber\\ &&\left[ \int_\chi^{\chi_H}{\rm d} \chi'
n(\chi') {f_K(\chi'-\chi)\over f_K(\chi')}\right]^2,
\label{pofkappa}
\end{eqnarray}
were $f_K(\chi)$ is the comoving angular diameter distance out to
radial distance $\chi(z)$, and $n(\chi)$ is the redshift distribution of
the sources. $P_{3D}\left({k\over f_K(\chi)},\chi\right)$ is the
3-dimensional mass power spectrum, and $\kappa$ is a 2-dimensional
wave vector perpendicular to the line-of-sight.

Cosmic shear can be studied using three different 2-point statistics,
which differ only by their filtering schemes. These various statistics
have different wavelength sensitivities to the power spectrum and
therefore the effect of systematics on each is different. This enables
the comparison of multiple cross-checked solutions. Two-point
statistics are measured as a function of scale $\theta_c$, which could
either be a galaxy pair separation or smoothing window radius. The
relation between each two-point statistics and the power spectrum of
the gravitational convergence ({\it i.e.} the projected dark matter
power spectrum) can be expressed as follows:

\begin{itemize}
\item Top-hat variance:
\begin{equation}
\langle\gamma^2\rangle_{\theta_c}={2\over \pi\theta_c^2} \int_0^\infty~{{\rm d}k\over k} P_\kappa(k)
[J_1(k\theta_c)]^2.
\label{theovariance}
\end{equation}

\item Shear correlation function:
\begin{equation}
\langle\xi\rangle_{\theta_c}={1\over 2\pi} \int_0^\infty~{\rm d}
k~
 k P_\kappa(k) J_0(k\theta_c).
\label{theogg}
\end{equation}
\item Aperture mass variance:
\begin{equation}
\langle M_{\rm ap}^2\rangle_{\theta_c}={288\over \pi\theta_c^4} \int_0^\infty~{{\rm d}k\over k^3}
P_\kappa(k) [J_4(k\theta_c)]^2 ,
\label{theomap}
\end{equation}
\end{itemize}
with the aperture mass variance  defined as:
\begin{equation}
M_{\rm ap}(\theta_c)=\int_{\theta < \theta_c}~{\rm d}^2\thetag \ \kappa(\thetag)~U(\theta),
\end{equation}
where $U(\theta)$ is a compensated filter such as:
\begin{equation}
U(\theta)={9\over \pi \theta_c^2} \left(1-{\theta^2\over\theta_c^2}\right)
\left({1\over 3}-{\theta^2\over\theta_c^2}\right).
\label{Ufilter}
\end{equation}

$M_{\rm ap}$ can be expressed in terms of the tangential shear
component inside a circle as follows (\cite{K94,S96}) :
\begin{equation}
M_{\rm ap}(\theta_c)=\int_{\theta < \theta_c}~{\rm d}^2\thetag \ \gamma_t(\thetag)~Q(\theta),
\label{mapfromshear}
\end{equation}
where the tangential shear component  $\gamma_t(\thetag)$ at the position
$\thetag$ is given by:
 \begin{equation}
\gamma_t(\thetag)=-{\mathcal Re}\,\left(\gamma\left(\thetag\right)\right)e^{-2i\phi}
\label{tangcomp}
\end{equation}
and the function $Q(\theta)$ is defined as :

\begin{equation}
Q(\theta)=\frac{2}{\theta^2}\int^\theta_0 d\theta'~\theta'~U(\theta')-U(\theta)\,.
\end{equation}

 The aperture mass statistic as a tool for the cosmic shear
analysis has been discussed in many papers
(\cite{SVWJK98,P02al,MC03,MV05}). This statistic is sensitive to
curl-free correlations (E-modes) generated by the (scalar)
gravitational potential.  Curl correlations (B-modes) are then easily
derived using the same statistics, after rotating each galaxy by
$45\,deg$.  If the only signal present is due to lensing, then the
B-modes should be zero at all scales. This simple procedure is
therefore a powerful diagnostic tool to assess systematic residuals in
cosmic shear signal.

Unfortunately, the $M_{ap}$ statistic is sensitive to the smallest
accessible angular scales, where cosmic shear signal depends on the
poorly-known non-linear evolution of the dark matter power
spectrum. This shortcoming forces us to compute E- and B-modes on
larger angular scales in a different way. For this we use the top-hat
shear variance and the shear correlation functions.  These functions
are usually derived from the $\xi_+$ and $\xi_-$ shear correlation
functions:

\begin{eqnarray}
\xi_+(r)&=&\langle \gamma_t(\theta)\gamma_t(\theta+r)\rangle+\langle \gamma_r(\theta)
\gamma_r(\theta+r)\rangle.\nonumber \\
\xi_-(r)&=&\langle \gamma_t(\theta)\gamma_t(\theta+r)\rangle -\langle
 \gamma_r(\theta)\gamma_r(\theta+r)\rangle,
\label{xipm}
\end{eqnarray}
where $\gamma_t$ and $\gamma_r$ are the tangential and radial
projections of the shear onto the local frame joining two galaxies
separated by a distance $r$.  Following \cite{CNPT01a}, we define

\begin{equation}
\xi'(r)=\xi_-(r)+4\int_r^\infty \frac{dr'}{r'} \xi_-(r')-12r^2 \int_r^\infty \frac{dr'}{r'^3}\xi_-(r').
\label{eqn:xipr}
\end{equation}
The $E$ and $B$ shear correlation functions are given by
\begin{equation}
\xi^E(r)=\frac{\xi_+(r)+\xi'(r)}{2}\ \ \ \ \ \
\xi^B(r)=\frac{\xi_+(r)-\xi'(r)}{2}.
\label{eqn:xieb}
\end{equation}

A similar relation can be found for the aperture mass and the top-hat
statistics as showed in \cite{CNPT01b}.  \cite{CNPT01b} also pointed
out $\xi^E$ and $\xi^B$ can only be derived up to an integration
constant which depends on the extrapolated signal outside the
measurement range.

Finally, the amplitude of the lensing signal depends on the galaxy
redshift distribution $n(z)$ (see Eq.(\ref{pofkappa})).  As in
previous works (see \cite{VWM02al,VWM05al}), we use the following
redshift distribution:
\begin{equation}
n(z)=\frac {\beta}{z_s\; \Gamma \Big(\frac{1+\alpha}{\beta}\Big)}
\Big(\frac{z}{z_s}\Big)^\alpha\;\exp\,\Big[{-\Big(\frac{z}{z_s}\Big)^\beta}\Big] \ ,
\label{eq1}
\end{equation}
where $\alpha$, $\beta$ and $z_s$ parameters are derived from deep
photometric redshift catalogues.  
The lensing signal can be predicted for any redshift range using Eq.(\ref{pofkappa}) and Eq.(\ref{eq1}).

\section{The Deep CFHTLS T0001 data set}

The Deep CFHTLS data used in this work consists of $u^*$, $g'$, $r'$,
$i'$ and $z'$ stacked {\sc Megacam} images that form the first CFHTLS
release (hereafter T0001).  The release is composed of stacked images,
catalogues and relevant meta-data produced from observations in four
uncorrelated fields that were carried out at CFHT with the {\sc
Megaprime} instrument between June 1st 2003 and July 22, 2004.
Details regarding each field are listed on the CFHTLS web
pages\footnote{ {\tt
http://www.cfht.hawaii.edu/Science/CFHLS/

cfhtlsdeepwidefields.html}}.

Each {\sc Megacam} image consists of an array of 9 $\times$ 4 EEV CCDs
of 2048 $\times$ 4612 pixels each (\cite{Boul03}).  The pixel scale is
0.186" and the camera covers a total field of 1 degree $\times $ 1
degree. There are two large gaps of 82 arc-second between rows of
CCDs. In order to produce complete fields, the gaps have been filled
by organizing observations in a series of exposure sequences with
large offsets. This results in an heterogeneous pixel illumination at
the borders of each CCD.  This spatial flux variation induces a
varying pixel signal-to-noise ratio that is taken into account by
using pixel weight maps together with hand-made masks (see Section 4)
to discard noisy areas of each field.

The stacks include only {\sc Megacam} images with seeing better than
1.0" \footnote{We use the seeing definition of {\sc Terapix} as twice
  the median flux radius of a selection of point sources on each CCD.
  {\sc Flux\_radius} as measured by {\tt SExtractor}, is the radius of
  the disk that contains 50\% of the total flux. For a Gaussian
  profile the {\tt SExtractor} seeing is almost equal to the Full
  Width at Half Maximum (FWHM). For a typical {\sc Megacam} PSF, it is
  slightly larger (10\%) than the true PSF FWHM.} and airmass below
1.4 have been selected. However, because there were fewer $u^*$-band
images than for the other filters, we relaxed the selection criteria
for this filter and kept all $u^*$ images with seeing below 1.4".
Only three of the four Deep fields have been selected for cosmic shear
studies.  The D2 Deep field has been dropped from our sample because
it is significantly shallower than the other three fields.

Data were calibrated and processed at CFHT and the {\sc Terapix} data
center. The full T0001 release is archived at CADC\footnote{{\tt
http://cadcwww.dao.nrc.ca/cfht/cfhtls/}} and available to any CFHTLS
registered user.  A description of the data processing pipeline used
to produce the deep T0001 stacks is beyond the scope of the paper, but
the details can be found on the {\sc Terapix} web pages\footnote{{\tt
http://terapix.iap.fr/article.php?id\_article=382}}.  Photometric and
astrometric methods and quality assessments done on these data are
explained in a short explanatory supplement\footnote{{\tt
http://terapix.iap.fr/article.php?id\_article=383}}.  The processing
(astrometric and photometric calibrations, pixel re sampling, image
warping and stacking, catalogue production) uses the current first
generation {\sc Terapix} software tools and closely follows the one
used for the {\sc Virmos}-Descart survey that is described in
\cite{HJMCC03}. We refer to this paper, and to the {\sc Terapix} and
CFHT\footnote{ {\tt
http://www.cfht.hawaii.edu/Science/CFHTLS-DATA/

dataprocessing.html }}
web pages for further details.

The accuracy of the photometric calibrations can be estimated from the
stellar colour-colour plots and the galaxy counts in all bands given
on the {\sc Terapix} T0001 pages and is also discussed in the more
detailed stellar analysis done by Schulteiss et al. (in
preparation). In all bands, the cumulative internal and systematic
photometric errors are 0.05 mag up to AB=22.5, and never larger than
0.1 to the 80\% completeness limit ($\approx$ AB=25.5).  This
uncertainty is sufficient for the cosmic shear studies on this paper.

\begin{table*}
\caption{Summary table of T0001 D1, D3 and D4 deep stacks used in this
work. Magnitudes are instrumental AB.  Details on magnitude, aperture,
seeing and completeness definitions are given in the explanatory page
{\tt http://terapix.iap.fr/article.php?id\_article=383}}
\label{tablestack}
\begin{center}
\begin{tabular}{|l|c|c|c|}
\hline
 & D1 & D3 & D4 \\
\hline
RA (J2000) &  02:25:59 &  14:19:27 &  22:15:31 \\
DEC (J2000) &  -04:29:40 & +52:40:56 & -17:43:56 \\
Effective FOV (deg$^2$) &  0.80 &  0.77 &  0.77 \\
\hline
Exp. time $u^*$ (s) &  10560 &  4620 &  16680 \\
Median seeing $u^*$ (arc-sec.) &  1.15 &  0.88 &  1.05 \\
Completeness $u^*$ 50\% (mag.) &  26.4 &  26.0 &  26.2 \\
\hline
Exp. time $g'$ &  7515 &  8010 &  11250 \\
Median seeing $g'$ (arc-sec.) &  0.98 &  0.95 &  0.99 \\
Completeness $g'$ 50\% (mag.) &  26.4 &  26.5 &  26.2 \\
\hline
Exp. time $r'$ &  17280 &  20820 &  26400 \\
Median seeing $r'$ (arc-sec.) &  0.87 &  0.93 &  0.85 \\
Completeness $r'$ 50\% (mag.) &  26.1 &  26.4 &  25.9 \\
\hline
Exp. time $i'$ &  52000 &  59640 &  58800 \\
Median seeing $i'$ (arc-sec.) &   0.88 &  0.92 &  0.88 \\
Completeness $i'$  50\% (mag.) &  26.1 &  26.2 &  25.8 \\
\hline
Exp. time $z'$ &  12240 &  15120 & - \\
Median seeing $z'$ (arc-sec.) &   0.86 &  0.85 & - \\
Completeness $z'$ 50\% (mag.) &  24.5 &  24.6 & - \\
\hline
\end{tabular}
\end{center}
\end{table*}

Table \ref{tablestack} summarizes the T0001 stacks used in this work.
The completeness limits have been computed by adding randomly
simulated stars (Moffat profiles) inside a 2000$\times$2000 area of
each Deep field and by running the detection and photometry again,
using the {\tt MAG\_AUTO} magnitude of {\tt SExtractor}\footnote{{\tt
http://terapix.iap.fr/rubrique.php?id\_rubrique=91}} software
(\cite{BA96}).  The completeness was also checked using
galaxy counts \footnote{see {\tt http://clix.iap.fr/T0001/Plots/

           CFHTLS\_D\_i\_galcount\_T0001.png} 

and {\tt http://clix.iap.fr/T0001/Plots/

           CFHTLS\_D\_r\_galcount\_T0001.png}}.

The D4-$z'$ data have not been released by {\sc Terapix} because the
astrometric internal accuracy was below the scientific
requirements. The large rms error found in the D4-$z'$ data is not
been fully understood, but it results in a large number of galaxy
mismatches during the $u^*,g',r',i',z'$ catalogue
cross-identification. This hampers reliable panchromatic studies for
many galaxies detected in this field. A further investigation reveals
that the D4$-i'$ astrometric solution is also slightly off, while data
quality in other filters are excellent. Although it has no impact on
the D4-$u^*,g',r',i'$ photometric studies, a quick weak lensing
analysis of the D4-$i'$ field shows it has more systematic residuals
than D1-$i'$ and D3-$i'$. In contrast, the three Deep $r'$ band data
have similar quality and do not show systematics residual differences.
We therefore used the $r'$ band as the reference data set for all
comparison between the fields, and only use the deep D1/D3 $i'$ band
data for colour comparisons, when needed.


\section{Detection  of the shear signal}
\subsection{Galaxy shape parameters}
Catalogues and shape measurements of galaxies are produced using the
{\tt IMCAT} software (\cite{KSB95}, hereafter KSB).  For each object
the centroid position and the half-light radius $r_h$ are
measured. These parameters are then used to derive orientations and
raw ellipticities of galaxies from the weighted second moments
$I_{ij}$ of the galaxy light distribution.  In order to minimize the
noise contribution each moment is filtered using a Gaussian filter
$W(\theta)$ of size $r_h$:

\begin{equation}
I_{ij}= \int d^2 \bd \theta \, W\left(\theta\right) \
 \theta_i \,\theta_j \, (\bd {\theta})\ f(\bd{\theta})\;,
\end{equation}
where $f(\bd{\theta})$ is the surface brightness.

The raw ellipticity is given by:
\begin{equation}
\bd{e}=\left(\frac{I_{11}-I_{22}}{{\mathcal Tr \left(I\right)}}; \; \frac{2\,I_{12}}{{\mathcal Tr\left(I\right)}}\right)\;.
\end{equation}
where $\mathcal Tr (I)$ represents the trace of the matrix $I$.  We
use the KSB method to get an unbiased estimator of the shear $\bd
\gamma$.  This method has been tested by several teams and it has been
demonstrated that it provides robust and reliable shear measurements
from ground based data (see the comprehensive critical investigation
of KSB and other techniques by \cite{Hey05b}, and also references
therein).

\subsubsection{PSF correction: the principle}

Let us assume the shear-free intrinsic ellipticity of a galaxy is
$\bd{e}^{0}$. On the detector, its shape is eventually modified by the
distortions produced by gravitational lensing effects and systematics
that increase the smearing and the anisotropic component of the PSF
(atmosphere, optical aberrations).  Assuming these distortions are
small, \cite{KSB95} demonstrated the observed ellipticity,
$\bd{e}^{obs}$, can be written:
\begin{equation}
\bd{e}^{obs}_\alpha=\bd{e}^{0}_\alpha+P^{sh}_{\alpha\beta}\bd{\gamma}+P^{sm}_{\alpha\beta}\bd{q} \ ,
\label{eqn:ell}
\end{equation}
where $\bd q$ is the anisotropic component of the PSF and $\bd \gamma$
is the gravitational shear.  \ $P^{sh}$ and $P^{sm}$ are called the
shear and the smear polarisability.  Their values depend on the galaxy
surface brightness and on the filter properties $W(\theta)$.  \ $\bd
q$ can be derived directly from the data, by measuring the ellipticity
of stars in each field, $\bd e_\star$, such as:
\begin{equation}
\bd{q}_\alpha=\frac{\bd{e^\star}_{\alpha}}{P^{sm}_{\beta\beta}}\;.
\end{equation}

The shear polarisability is however altered by the isotropic smearing
component of the PSF.  It results in a modification of the shear
polarisability
\begin{equation}
P^{\gamma}=P^{sh}-\frac{P^{sh}_{\star}}{P^{sm}_{\star}}P^{sm}\;,
\end{equation}
where $P^{\gamma}$ is called pre-seeing shear polarisability and
$P^{sh/sm}_{\star}$ refers to stars (\cite{LK97}).  Provided the
assumption $\langle e_{0}\rangle = 0 $ is valid, an unbiased estimator
of the shear $\bd{\gamma}$ is given by:
\begin{equation}
\bd{\gamma}=\langle P_{\gamma}^{-1}(\bd{e}^{obs}-P^{sm}\bd{q}) \rangle.
\label{extgamma}
\end{equation}

\subsubsection{Object selection}

Prior to cosmic shear analysis, all CFHTLS images are checked by eye
and masks are drawn by hand.  These masks are designed to avoid
elongated defects, like saturated stars, as well as large foreground
galaxies with extended bright halo that may contaminate the shape of
underlying faint galaxies (see \cite{VWM01} for details). We should
emphasize that masks are {\it only} drawn using criteria ($z=0$
galaxies, bright stars, CCD defects) that are not correlated with the
lensing signal.  In addition, we used the weight map images produced
by {\sc Terapix} for each stack to reject all pixels with a relative
weight amplitude less than of $80\%$. This rejection step reduces
significant spatial variation of the detection threshold and keeps the
averaged redshift distribution of lensed galaxies stable over the
field.  The rejection scheme removes the CCD boundaries from all of
the fields, and is essentially equivalent to singling out each CCD
region, as was done earlier in {\sc Virmos}-Descart survey.  The gain
in homogeneity is however preserved at the expense of the sky
coverage. About $30\%$ of the initial area is lost after the masking
process.

Stars needed for the PSF correction are selected along the stellar
locus of the magnitude/size diagram (\cite{Fah94}), from the region
where stars are about one magnitude fainter than the saturation level
and where they cannot be confused with faint galaxies. The
$P^{sm}_{\star}$ and $P^{sh}_{\star}$ values are derived at all {\sc
Megacam} image positions from a PSF mapping that samples the PSF
smearing and PSF anisotropy at the position of each star, and by
interpolating their values between the stars.  This operation is done
on each CCD separately, as suggested by \cite{H04}.  The PSF is mapped
using a composite model of a second order polynomial and a rational
function, $p^\alpha(x,y)$:
\begin{eqnarray}
&&p^\alpha(x,y)=  \\
&&a_0+a_1x+a_2y+a_3x^2+a_4xy+a_5y^2+ c(x,y)\nonumber
\end{eqnarray}
where $c(x,y)$ is the rational function chosen as:
\begin{eqnarray}
&&c(x,y)=\\
&&\frac{b_0+b_1x+b_2y+b_3x^2+b_4xy+b_5y^2+b_6y^3+b_7y^4}{1+b_8x+b_9y}\nonumber
\end{eqnarray}
The second order polynomial terms models the smooth low frequency PSF
component, while the rational function provides a model for the high
frequency PSF terms (\cite{H04}).

The correction is made in two steps. First, the coefficients of the
rational function are determined. Since the CFHTLS Deep fields are
much deeper than the RCS (\cite{HOEK02}) and the {\sc Virmos}-Descart
(\cite{VW00,VWM01}) surveys, the density of selected stars is higher
and we do not need to map the PSF using external stellar fields.  Each
field has about 100 stars per CCD, so the high frequency PSF terms can
be reasonably well sampled down to 0.5 arc-minute, and all
coefficients of the rational function can be constrained with
sufficient accuracy.  In a second step, the polynomial terms are
determined.

We also compared the rational function solution against the second
order polynomial interpolation. We found the results are not very
different from our composite model, although the rational function
improves the quality and stability of the PSF mapping.

Once ellipticity is corrected we keep in the sample all objects with
angular size larger than the seeing disk and smaller than two
arc-seconds.  Following \cite{VW00}, we automatically reject one
galaxy in every close pairs with angular separation less than
$12\,arcsec$ in order to avoid contamination of ellipticity
measurements by overlapping isophotes of neighboring galaxies.

\begin{figure}
\begin{center}
\includegraphics[width=8cm,height=7cm]{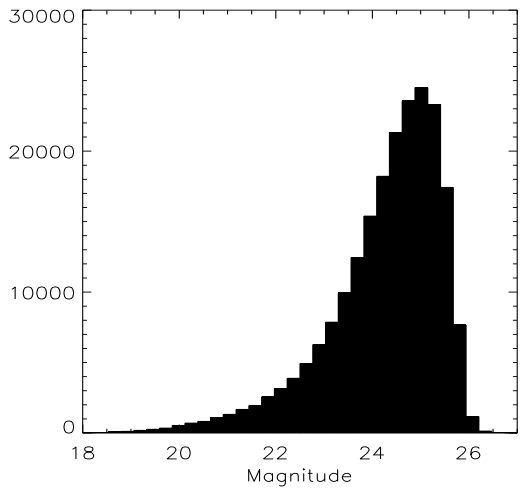}
\caption{\label{histomag} Magnitude distribution of galaxies in the three fields in the $r$
band.  More detailed galaxy count plots, for each filter and for each
Deep field, are available on the web at {\tt
http://terapix.iap.fr/article.php?id\_article=382}}
\end{center}
\end{figure}

The magnitude distribution of the final object catalogue is shown in
Fig.\ref{histomag}. The limiting magnitude corresponding to a 80\%
completeness limit is $r'_{AB}=25.5$.
Bright objects with magnitude smaller than $21.5$ and faint objects
with magnitude larger than $25.5$ are also removed from the galaxy
sample.  The final galaxy number density of the cosmic shear catalogue
is about $20 / arcmin^2 $.

As proposed by \cite{E01} we assign an ellipticity dispersion
$\sigma_g$ to each object corresponding to the ellipticity dispersion
in a box containing its 20 nearest neighbors in the
($magnitude,\,size$) space.  Weighted 2-point statistics are computed
assigning to each galaxy a weight given by $1/(\sigma_g^2+\sigma_e^2)$
where $\sigma_e$ is the ellipticity dispersion of the unlensed
galaxies.  A different noise estimation (\cite{HOEK00}) gives similar
results.

\begin{figure*}[!t]
\begin{center}
\includegraphics[width=5.5cm,height=5.5cm]{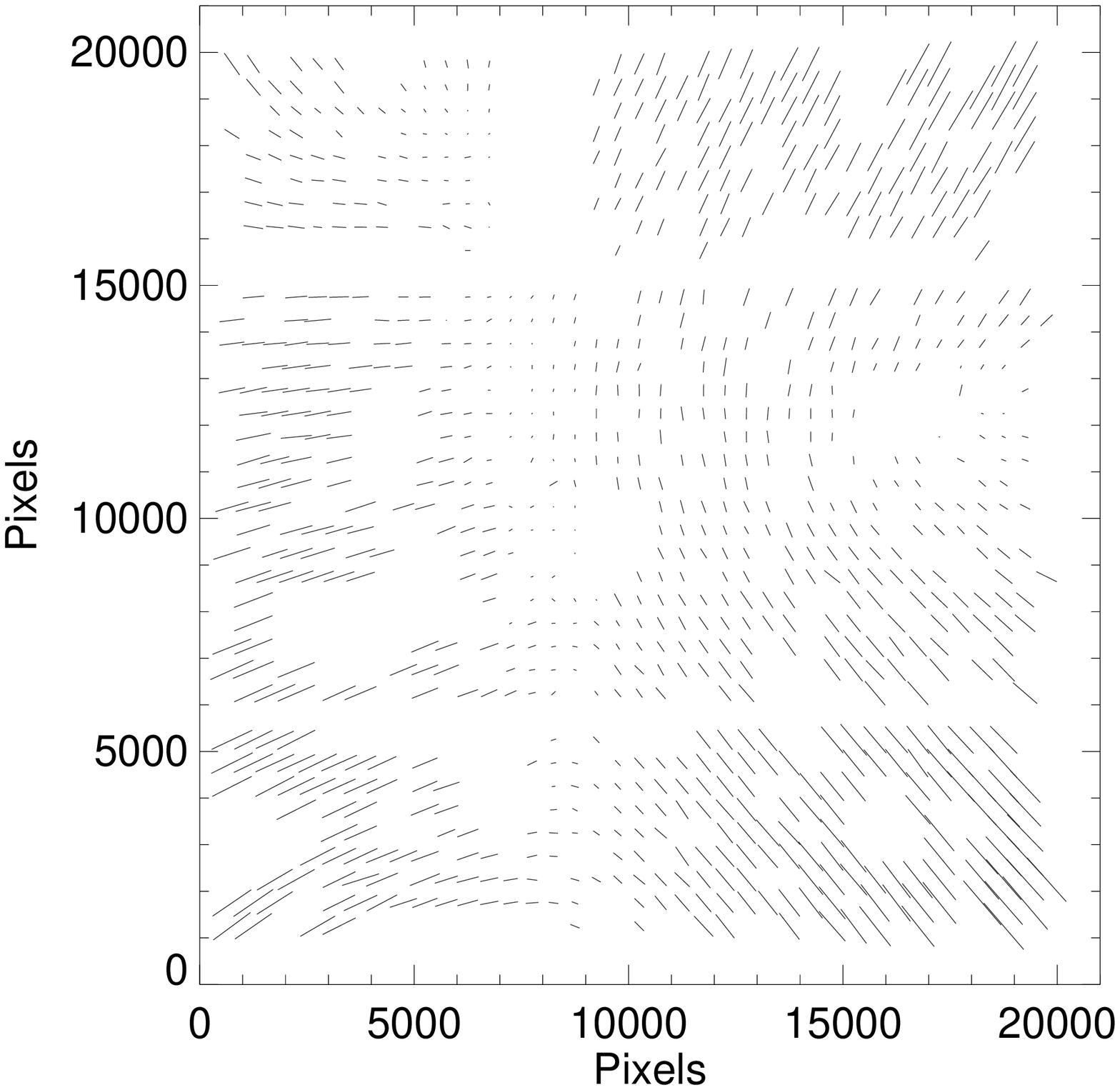}
\includegraphics[width=5.5cm,height=5.5cm]{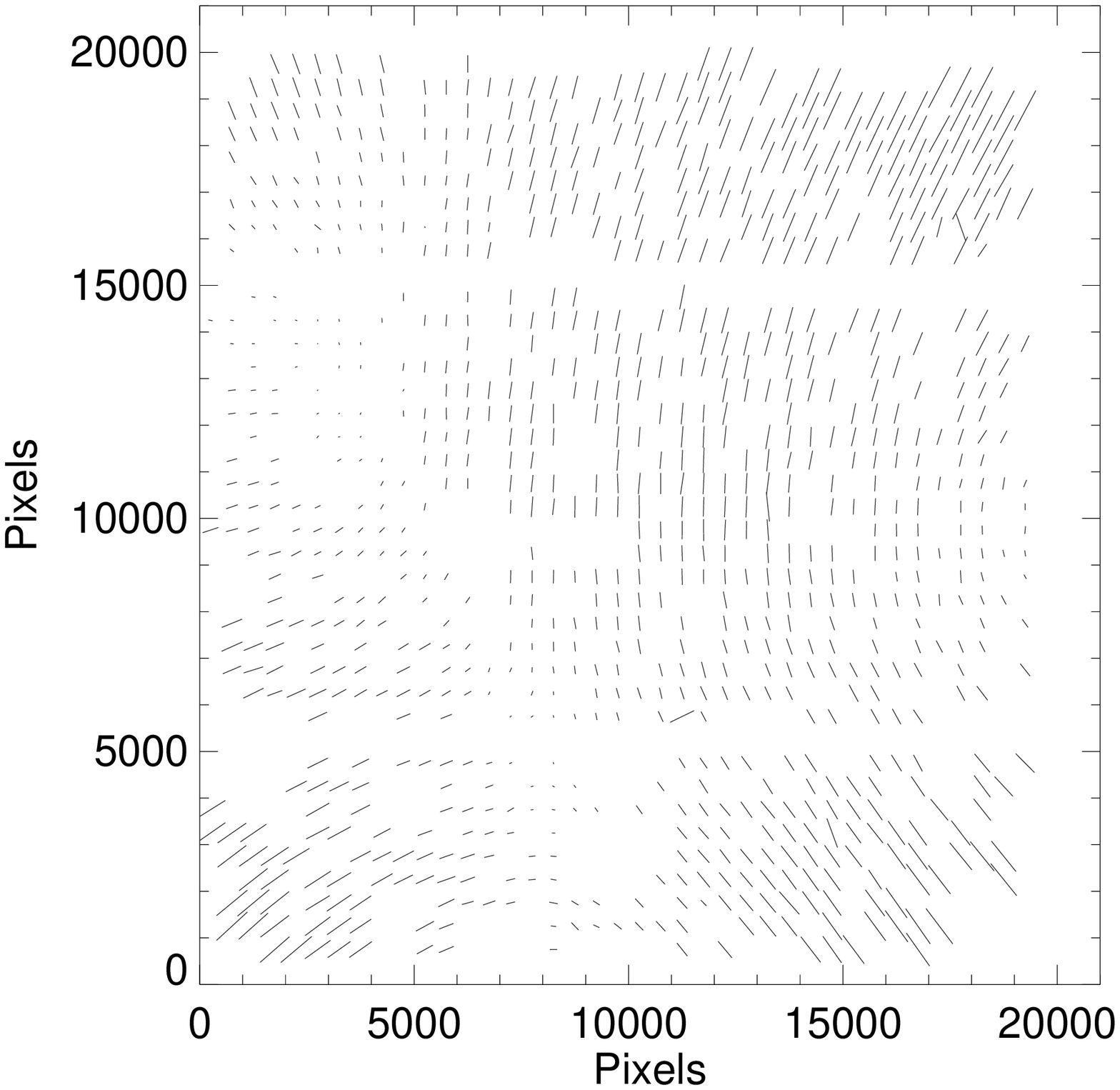}
\includegraphics[width=5.5cm,height=5.5cm]{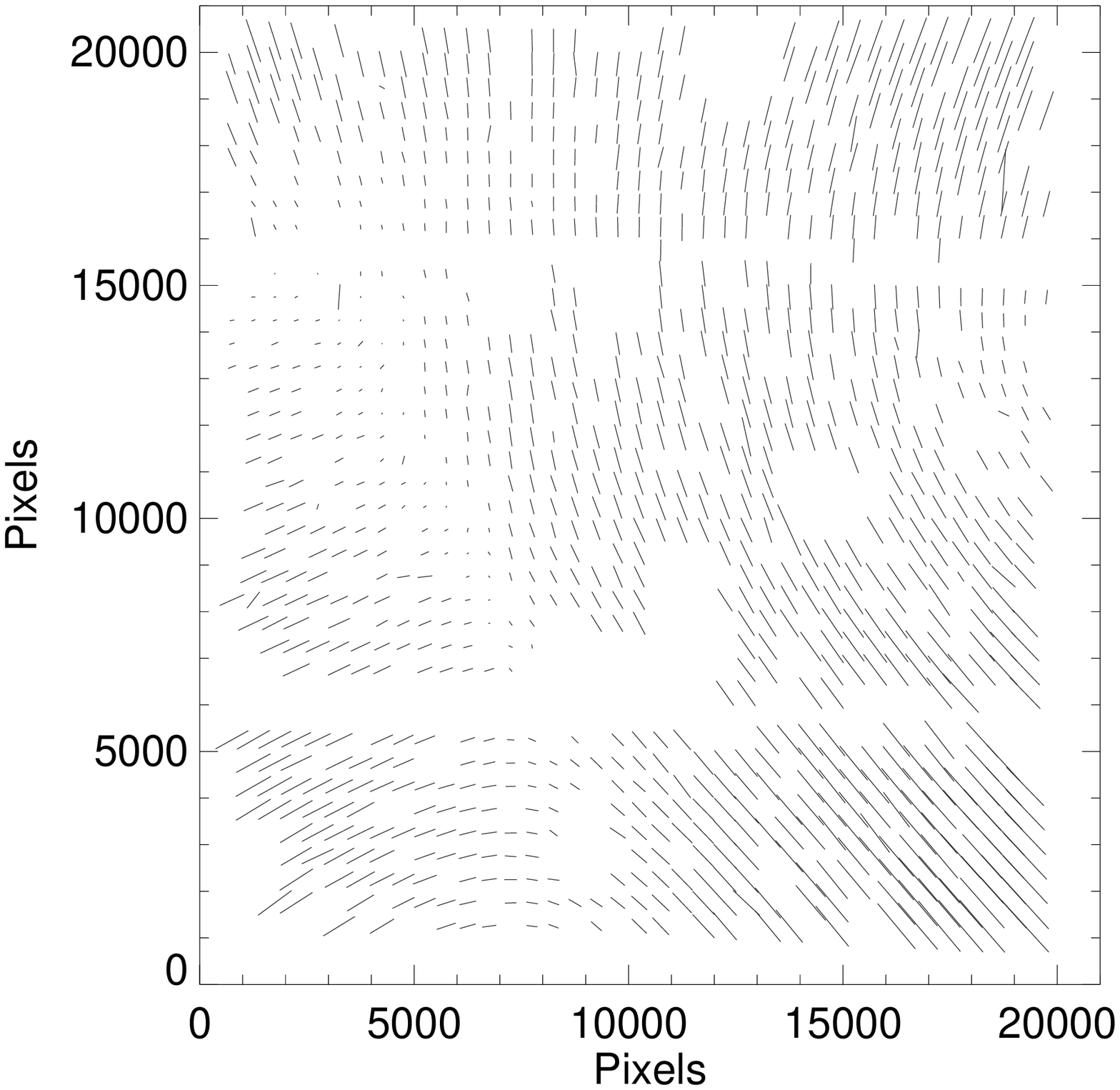}
\caption{\label{psfcarte} The mean  ellipticity of the stars shows the behavior of the  PSF anisotropy  for the three fields D1, D3 and D4 . For all of  the three fields, $<e>$ is few percent in the central part and it becomes about $10\%$ in the corners (see also fig.\ref{psf}).}
\end{center}
\end{figure*}

\section{Residual systematics}

\subsection{Quality of the PSF correction}

A visual inspection of the {\sc MegaPrime} PSF (Fig.\ref{psfcarte})
shows that the PSF anisotropy has significant variation over the field
and may also be very large at the boundaries\footnote{This strong PSF
anisotropy has been considerably reduced by the CFHT staff, after the
T0001 release. It should no longer be a critical issue for next
releases.}.  The PSF correction is therefore a critical step and its
reliability demands careful verifications. In addition to the usual
B-mode analysis shown in the next section, in this section we carry
out several analyses of the systematics. 

\begin{figure}
\begin{center}
\includegraphics[width=9.5cm,height=8cm]{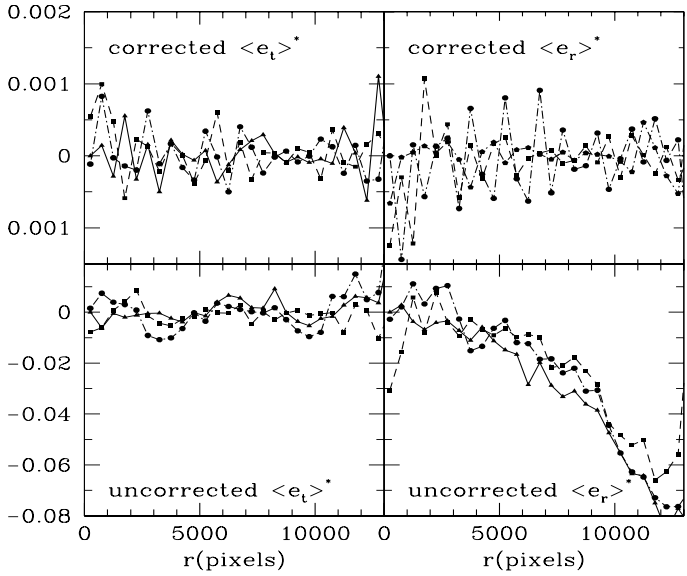}
\caption{\label{psf} Bottom panels show the mean tangential (left
panel) and radial (right panel) uncorrected stellar ellipticity as a
function of the distance to the center of the camera for D1rT001
(filled triangles ), D3rT001 (filled squares) and D4rT001 (filled
circles) fields. Top panels show the same quantities after PSF
correction.}
\end{center}
\end{figure}

The quality of the PSF correction and its homogeneity over the {\sc
Megacam} field camera can be assessed by comparing the mean star
ellipticity before and after PSF correction (Fig.\ref{psf}). The
average stellar ellipticities $\langle e_t\rangle$ and $\langle e_r
\rangle$ are plotted as a function of the radial distance from the
center of the field, $r$. It is interesting to note that the radial
ellipticity component degrades much more and much faster than the
tangential ellipticity.  However, the PSF correction done by the PSF
mapping is very good, for each Deep field. After correction, the
dispersion of star ellipticities is about 2 $\times$ 10$^{-3}$ at any
point of the camera. There is no significant change in the residual
error as function of position. The small increase in the fluctuation
of star shapes at very small distances is due to higher Poisson noise:
each radial bin has the same width, so the innermost circle
encompasses the smallest area and contains fewer stars than the
others.

\begin{figure}
\begin{center}
\includegraphics[width=9cm,height=7cm]{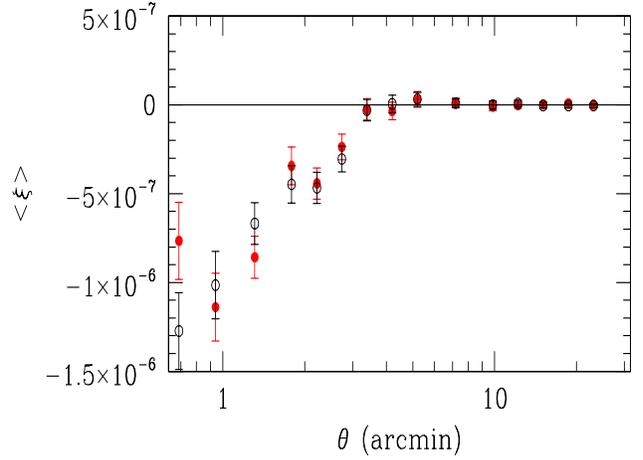}
\caption{\label{starspsf} E-modes (red filled circles ) and B-modes
(black open circles) top-hat two point statistics of corrected stars
show the smallness of residual PSF systematics.}
\end{center}
\end{figure}

Finally, we checked the residual amplitude of the shape correlation
function between corrected stars.  We found is to be two order of magnitude
smaller than the expected lensing signal at all scales probed by this
work (Fig. \ref{starspsf}).

\begin{figure}
\begin{center}
\includegraphics[width=8.5cm,height=12cm]{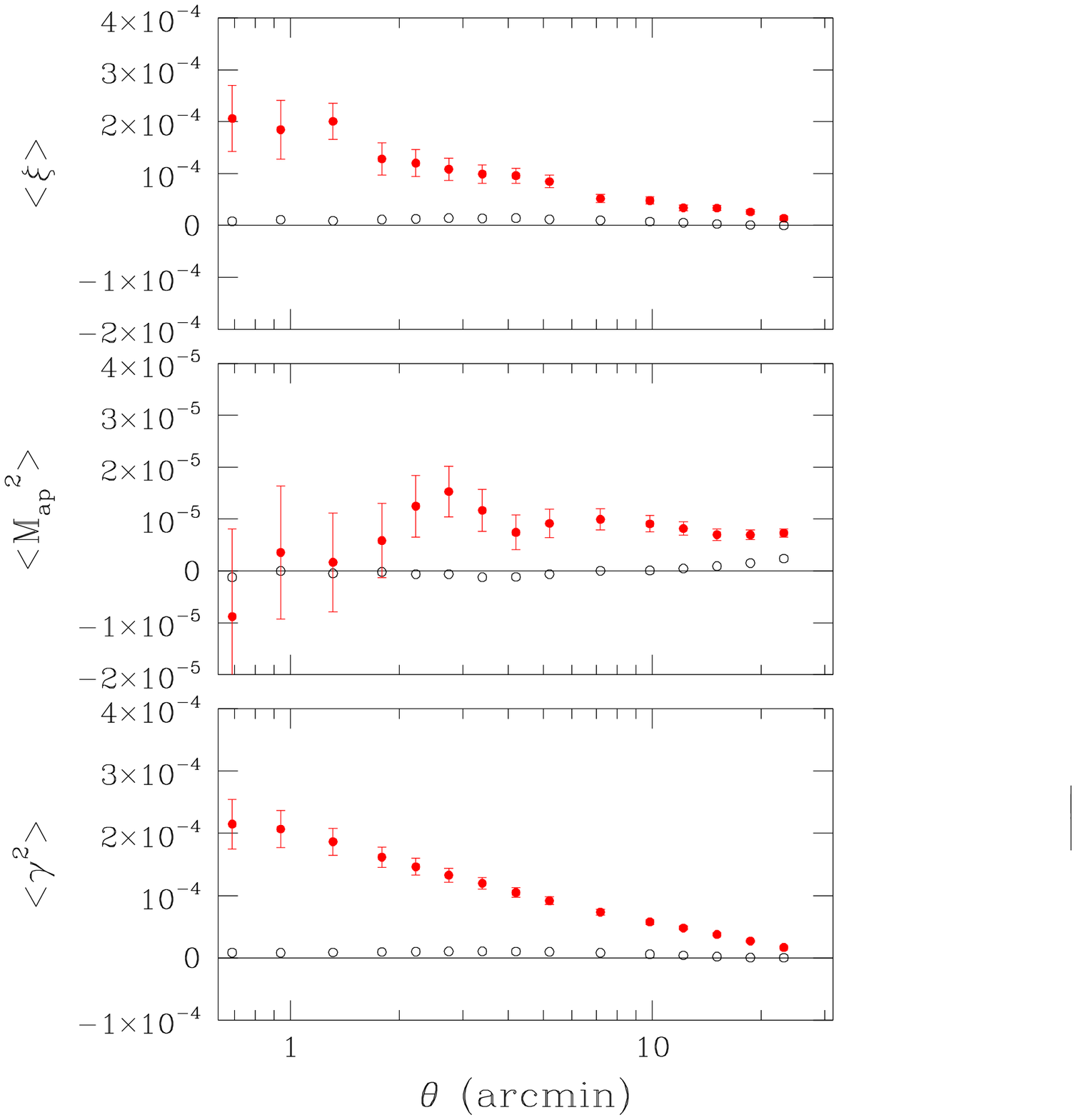}
\caption{\label{res1} Two-point signal statistics(red filled
circles):top-hat (bottom) and $M_{ap}$ (middle) correlation function
(top), compared with residual systematics (black open circles). Signal
error bars are statistical ones.}
\end{center}
\end{figure}

The tests discussed above only guarantee that the PSF correction is
excellent in the neighborhood of selected stars or on angular scales
larger than, or close to, the mean angular distance between stars.  In
regions where no stars were selected or on small scales, the local PSF
correction residuals may be larger than the average.  A useful test of
systematic residuals on small scales has been proposed by \cite{BRE03}
and \cite{Hey05b}.  Assuming the PSF model derived from stars and
applied to galaxies is unable to remove all systematic contributions,
the star-galaxy cross correlation will be non-zero and may vary as
function of angular scale.  If the residual is small, \cite{BRE03}
showed the systematic residual can be expressed as follows:
\begin{equation}
\xi_{sys}=\frac{\langle e_*\; e_{gal}\rangle^2}{\langle
e_*\;e_*\rangle} \ ; \label{sys}
\end{equation}
where $e_{gal}$ is the {\it corrected} galaxy ellipticity and $e_*$ is
the {\it uncorrected} star ellipticity.  We use the $\xi_{sys}$ to
compute the contribution of systematics for both top-hat and
compensated filter.  Fig. \ref{res1} shows they are consistent with
zero at all scales between 0.5 arc-minute to 30 arc-minutes.  This
confirms that residual systematics are negligible in the {\sc Megacam}
Deep fields.

\subsection{Independent analysis of $r'$ and $i'$ data}
The robustness of cosmic shear signal can also be assessed by
comparing results obtained using different filters for the same galaxy
sample.  Because gravitational lensing is achromatic, we expect the
shape and amplitude of cosmic shear to be identical for data taken in
different filters. Any significant difference between two bands
provides a diagnostic of the PSF corrections. A first attempt at
comparing shear measurements in different filters was made by
\cite{K00} using the CFHT12K camera. The $I$ and $V$ bands showed
significantly different signals that were inconsistent with the change
in redshift distribution between the two filters.

The Deep photometry provides a sample of the same galaxies detected in
different filters, so the signal is expected to be the same. However,
these filters have different depths, and the shallowest colours do not
have enough galaxies to allow a comparison of the signal between all
colours using the same galaxies. This limitation affects mainly the
$u^*$ and $g'$ bands.  Furthermore, these bands are more sensitive to
atmospheric dispersion than other filters.  We expect their PSF
anisotropy to be larger than for $r'$, $i'$ and $z'$ bands and its
correction may also depend more on the relative differences between
the averaged spectral energy distributions of stars used for the PSF
calibration and of galaxies. Hence, $u^*$ and $g'$ are not well suited
for weak lensing analysis.  The comparison between the $r'$ and the $z'$ bands  doesn't give many informations  because of the limited size of the matched sample. We therefore decided only to focus on the comparison between $r'$ and $i'$ bands. 

We computed the two-point statistics using the same objects in
$i'$-band and $r'$-band in the D1 and D3 fields only.  As reported
before, D4 was discarded from this study because it shows higher
systematic residuals in $i'$ bands than the two other fields.  It is
worth noting that both $r'$ and $i'$ band images have been processed
(flat fielding, astrometric and photometric calibrations, image
selection, image stacking) in a totally independent way.  The only
correlations between the two samples are the software tools and the
pipeline scheme used at {\sc Terapix}.

\begin{figure}[!ht]
\begin{center}
\includegraphics[width=9.5cm,height=6.5cm]{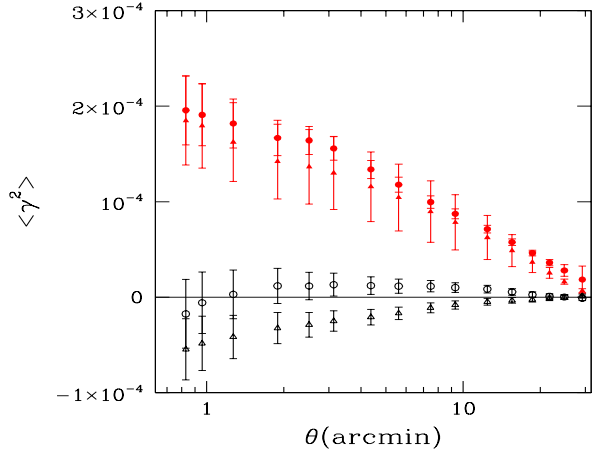}
\caption{\label{filtre}Top-hat variance of the E-modes in the i'-band
(red filled circles) and in the r'-band (red filled triangles) for the
same data set. Top-hat variance B-modes in the i'-band (black open
circles) and in r'-band (black open triangles).}
\end{center}
\end{figure}

The $r'$ and $i'$ ellipticity catalogues have been computed and
PSF-corrected independently, starting from the $r'$ and $i'$ T0001
stacked images. The galaxy cross-identification is done at the very
end of the processing to compare the results.  Fig. \ref{filtre} shows
the comparison of the $E$- and $B$-modes of the top-hat shear variance
for both the $i'$ and $r'$ data sets.  The error bars are estimated as
the quadrature sum of the statistical and the systematic error
$\xi_{sys}$ defined by Eq.(\ref{sys}). The amplitude of the latter is
bigger in the $r'$ band as shown by the residual B-modes in this
filter. The $r'$ and $i'$ bands results are remarkably similar, both
in shape and amplitude, they agree to within $1 \, \sigma$ at all
scales.

\section{Characterization of the shear signal}

\subsection{Two point statistics}
\begin{figure}[!ht]
\begin{center}
\includegraphics[width=8.5cm,height=12cm]{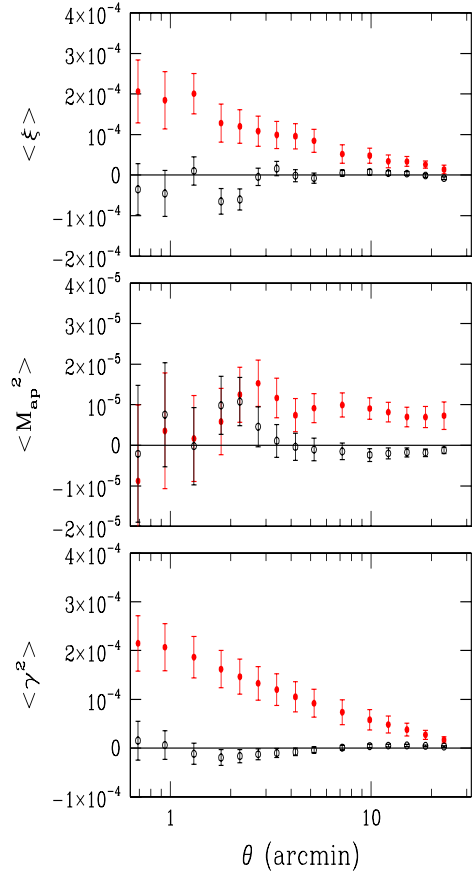}
\caption{\label{res2} Two-point statistics for all the three fields
combined. Red filled circles show E-modes, black open circles show
B-modes.  E-mode error bars include the statistical error and the
cosmic variance contribution, while B-modes are affected only by
statistical error.}
\end{center}
\end{figure}

The ellipticity correlation functions $\xi_+(r)$ and $\xi_-(r)$ are
measured from the weighted mean of all pairs with angular separation
$r$. The correlation function is computed using equations
(\ref{eqn:xipr}) and (\ref{eqn:xieb}).  The $M_{ap}$ and the top-hat
statistics are also computed as a function of the correlation
functions $\xi_+(r)$ and $\xi_-(r)$ following \cite{S02} and
\cite{CNPT01b}. Fig.\ref{res2} shows the two-point statistics for the
three deep fields D1, D3 and D4.  Error bars including statistical
noise and cosmic variance are computed from the $\xi_+(r)$ and
$\xi_-(r)$ as described in \cite{S02}.

The cosmic variance contribution is computed using the CFHTLS T0001
Deep survey properties: an effective density (after masking) of $20 \,
gal/arcmin^2$, an effective area of $2.1\,deg^2$, and an ellipticity dispersion per
ellipticity component of $0.3$ (the latter was measured from the
corrected ellipticity).  However, the error calculation described in
\cite{S02} is only valid for a single connected field with a number
density of $n$ equally-sized galaxies. We therefore replace the
statistical error component by the Poisson noise measured from the
data, using the weights (computed as described above) and positions of
each galaxy.  For the top-hat variance and the correlation function,
the free integration constant is chosen so that the B-modes on scales
between $15$ and $25$ arc-minutes vanish. Although its amplitude is
meaningless, one can see that the B-mode is flat and stable over that
range of angular scales.

In contrast, the $M_{ap}$ statistic does not have an undetermined
integration constant (as explained previously), so the B-mode
amplitude is a physical property.  Fig.\ref{res2} shows the presence
of B-modes. Note that the $M_{ap}$ filter for a given size
$\theta$ is mostly sensitive to scales around $ \simeq\theta/5$. This
explains why the other two-point statistics do not show B-modes at the
same scales. The B-mode at such small scales may result from intrinsic
alignment of galaxies (\cite{KS02,HH03}) or from the correlation
between intrinsic ellipticity and shear (\cite{HS04}). If these
systematics are real, we expect to correct them in future work by
using the photometric redshifts. A further investigation confirms
that the B-modes come from weak objects (i.e. $25.0 < r'_{AB} < 25.5
$), and that a magnitude cut that rejects objects with magnitude
fainter than $25.0$ gives zero B-modes at all the scales, even for the
$M_{ap}$ statistic.  However, we keep these objects in our catalogues
because a deep sample will be necessary to study the evolution of
signal with redshift. In addition, the presence of B-modes at small
scales will taken into account when we estimate cosmological
parameters.


\subsection{Evolution of signal with redshift}

The cosmological nature of the two-point statistical signal can be
established by comparing its amplitude as function of source redshifts
with theoretical expectations of the gravitational instability
paradigm and the gravitational lensing theory (\cite{B97, JS97}). To
first order, the signal should increase as $ z_s^{1.5}$ (\cite{B97,
JS97}, so even a rough separation of galaxies into low- and
high-redshift populations should split the cosmological lensing signal
accordingly.

The CFHTLS T0001 data sets are well suited for this analysis. The
observations can be used to sample the high redshift universe up to $z
\simeq 1$.  There are enough of galaxies to divide into two subsets
based on their estimated photometric redshifts.

Photometric redshifts were measured using the {\tt hyper-z} public
software\footnote{{\tt http://webast.ast.obs-mip.fr/hyperz/}}
(\cite{Bolzo00al}).  {\tt hyper-z} uses the multi-band photometric
data of a galaxy to derive its most likely redshift and spectral
energy distribution (SED) based on the Bruzual \& Charlot evolution
models (\cite{Bru93}).

We used the D1 and D3 $u^*$, $g'$, $r'$, $i'$ and $z'$ images and the
D4 $u^*$, $g'$, $r'$, $i'$ images (the D4-$z'$ stacked image is
missing in T0001).  Photometric catalogues were produced by the {\tt
SExtractor} software. All galaxies were first detected in the $r'$
band reference image. Magnitude and colours of galaxies are then
computed using the $r'-$center positions and inside an aperture scaled
according to the size of each galaxy in $r'$-band. The $\chi^2$
minimization was performed assuming magnitude errors derived from {\tt
SExtractor}, which range between $\Delta mag =0.03$ and $\Delta mag
=0.1 $ in all bands.

\begin{figure}[!ht]
\begin{center}
\includegraphics[width=9.cm,height=7.cm]{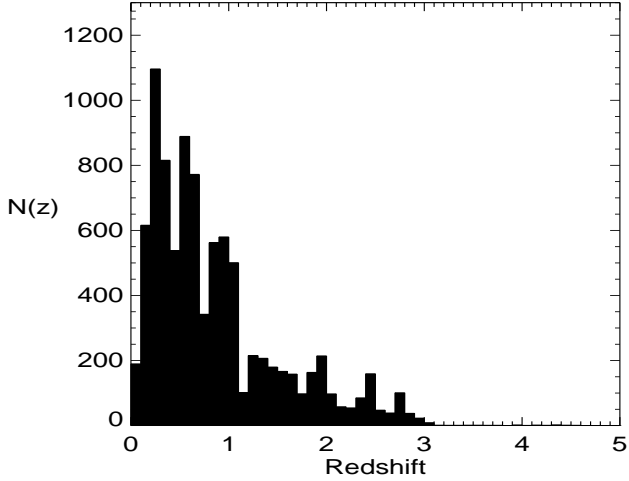}
\caption{\label{histoz}Density of galaxies of D1 field as a function
of photometric redshift.}
\end{center}
\end{figure}

Fig. \ref{histoz} shows the photometric redshift distribution of the
galaxies in D1 field down to $i'=24.0$. This subsample can be compared
with the VVDS spectroscopic redshift distribution obtained from 11000
spectra in the same region (\cite{Olf04}). There are no apparent
discrepancies that would make the separation into photometric low- and
high-redshift galaxies unreliable. Beyond $i'=24.0$, large
spectroscopic redshift samples are not yet available, but we don't
have any reason to believe that our photometric redshift accuracies
will degrade significantly for the $i'<24.0$ sample.

\begin{table}
\caption{Mean photometric redshift in magnitude bins.}
\label{tablemagz}
\begin{center}
\begin{tabular}{|l|c|}
\hline
magnitude bin & mean redshift \\
\hline
  $18.5 \leq i_{AB} \leq 24$ & $<z>=0.850$\\
  $19.0 \leq i_{AB} \leq 24$ & $<z>=0.853$\\
  $19.5 \leq i_{AB} \leq 24$ & $<z>=0.858$ \\
  $20.0 \leq i_{AB} \leq 24$ & $<z>=0.865$\\
  $20.5 \leq i_{AB} \leq 24$ & $<z>=0.876$\\
  $21.0 \leq i_{AB} \leq 24$ & $<z>=0.892$\\
  $21.5 \leq i_{AB} \leq 24$ & $<z>=0.913$\\
  $22.0 \leq i_{AB} \leq 24$ & $<z>=0.942$\\
  $22.5 \leq i_{AB} \leq 24$ & $<z>=0.981$\\
  $23.0 \leq i_{AB} \leq 24$ & $<z>=1.035$\\
  $23.5 \leq i_{AB} \leq 24$ & $<z>=1.100$\\
\hline
\end{tabular}
\end{center}
\end{table}

The cosmic shear catalogue can therefore be split into two samples
with equal numbers of galaxies at high and low redshifts with
reasonable confidence and can be compared with cosmological
predictions.  Poisson noise is therefore similar in the two
subsamples, but photometric redshift errors are expected to be larger
in the high-redshift tail.

\begin{figure}[!ht]
\begin{center}
\includegraphics[width=9.cm,height=7.cm]{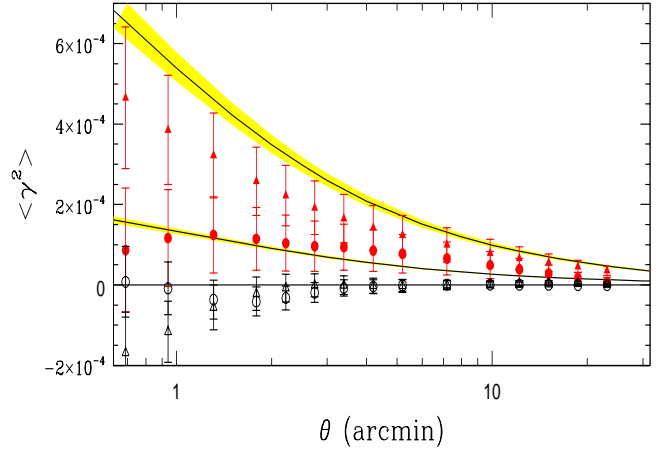}
\caption{\label{zphoto} Top-hat variance for ``high-$z$'' subsample
(red filled triangles) and ``low-$z$'' subsample'' (red filled
circles). B-modes for the two subsamples are also shown. Error bars
include statistical noise and cosmic variance.  The data are compared
with theoretical fiducial model ($\Omega_m=0.3$, $\Omega_\Lambda=0.7$,
$\sigma_8=0.88$ and $h=0.7$ ) and source distribution $n(z)$ modeled
by Eq.\ref{eq1}, with $\alpha=1.98$, $\beta=0.66$ and $z_s=0.0981$.  The
low-$z$ source selection is simulated using $n(z)$ between $0.3 < z <
1.2 $ and zero otherwise (bottom line).  Likewise, the red filled
triangles and the black open triangles are the E- and B-modes of the
``high-$z$'' sample. The data are compared with the same theoretical
model with a high-$z$ source selection simulated using $n(z)$ between
$ z > 0.8$ and zero otherwise.  Shaded areas show models within
$z_s=0.0981^{+0.013}_{-0.011}$ that represent the $1 \sigma$ error
region on $z_s$ as derived from the likelihood parameter estimation.}
\end{center}
\end{figure}

Fig. \ref{zphoto} shows the top-hat shear variance measured for the
two populations.  The low-$z$ sample ranges in $0.3 \lsim z \lsim 1.$,
while the high-$z$ galaxies have $z \gsim 1.0$. Error bars include
Poisson noise and cosmic variance (see Sect. 4.2).  The difference
between the two samples demonstrates the cosmological nature of the
signal.  An indicative comparison of signals with theoretical
predictions is also plotted.

The relative lensing amplitude for the two source galaxy populations
is less sensitive to cosmic variance fluctuations, and agrees with the
predictions.

Contamination by galaxies with incorrect photometric redshifts is
likely important, in particular for the faintest galaxies and the
high-$z$ tail ( further informations about degeneracy of photometric
redshifts in the case of missing infrared bands can be found on {\tt
hyper-z} user's guide).

In spite of potential contamination by incorrect photometric
redshifts, the cosmological imprint of large-scale structure detected
in the Deep CFHTLS data shows that {\sc MegaPrime} is suitable for
cosmic shear studies. Fig. \ref{zphoto} also demonstrates that
the CFHTLS Deep survey has promising potential for tomographic studies
that explore the evolution of the dark matter power spectrum with
look-back time.


\section{Parameter estimation}

\subsection{Derivation of the likelihood function}

In this Section, we describe the estimation of cosmological
parameters.  In a subsequent paper, we will perform a complete
parameter estimate, combining weak lensing with other cosmological
probes.  Therefore, here we limit the analysis to the normalization of
the mass power spectrum ($\sigma_8$) and matter density ($\Omega_m$)
measurements. The shape parameter $\Gamma$ is given by the Cold Dark
Matter paradigm $\Gamma=\Omega_m~h$, where $h$ is the reduced Hubble
constant.   
We allow the characteristic redshift of the source distribution to
vary around the best fit that will be described in the next
sub-section.

To measure cosmological parameters, we adopt a maximum-likelihood
method.  Let $d_i$ be the input data vector (i.e. the top-hat shear
variance as a function of scale $\theta_i$), and
$m_i(\Omega_m,\sigma_8,n(z))$ the prediction, function of the
parameters to be estimated. The likelihood function of the data is
then:
\begin{equation}
{\cal L}={1\over (2\pi)^n|\Cg|^{1/2}}
\exp\left[(d_i-m_i)\Cg^{-1}(d_i-m_i)^T\right],
\end{equation}
where $n = 16$ is the number of angular scale bins and $\Cg$ is the
$16\times 16$ covariance matrix of the top-hat shear,
\begin{equation}
C_{ij}=\langle (d_i-m_i)^T(d_j-m_j)\rangle,
\end{equation}
and $\Cg$ can be decomposed as $\Cg=\Cg_n+\Cg_s$, where $\Cg_n$ is the
statistical noise and $\Cg_s$ the cosmic variance covariance matrix.

As discussed above, the matrix $\Cg_s$ is computed according to
\cite{S02}, assuming an effective survey area of the CFHTLS Deep
fields: $2.1$ square degrees, a number density of galaxies $n_{gal}=20
/arcmin^2$, and an intrinsic ellipticity dispersion of $\sigma_e=0.3$
per component. 

The cosmic variance is computed assuming Gaussian statistics. While
this assumption becomes inappropriate on small angular scales, errors
on such scales are dominated by the statistical noise contribution, so
the Gaussian approximation remains an excellent one (\cite{VWM02al}).
The covariance matrix components are derived for a fiducial
cosmological model corresponding to the best fit of WMAP data proposed
by \cite{Sp03al}: $\Omega_M=0.3$, $\Omega_\Lambda=0.7$,
$\sigma_8=0.88$, $\Gamma=0.21$ (the reduced Hubble constant is
$h$=0.7). The $B$-mode is calibrated by marginalizing around $B=0$
within the 1$\sigma$ interval.

\subsection{Parameter estimation}

The source redshift distribution is calibrated using the Hubble Deep
Field (HDF) catalogues (\cite{SL98}), which provide a more accurate
estimate of redshift in absence of infrared data in CFHTLS fields. It
turns out that the F606 filter of WFPC2 is a good match to the {\sc
Megacam} r' filter within our $1\,\sigma$ magnitude error. We select all galaxies with $21.5 < r'<
25.5$. The Hubble Deep Fields provide a sample at
high redshifts that overlaps with the redshift range expected for the
CFHTLS Deep fields.

\begin{figure}[!ht]
\begin{center}
\includegraphics[width=9cm,height=7cm]{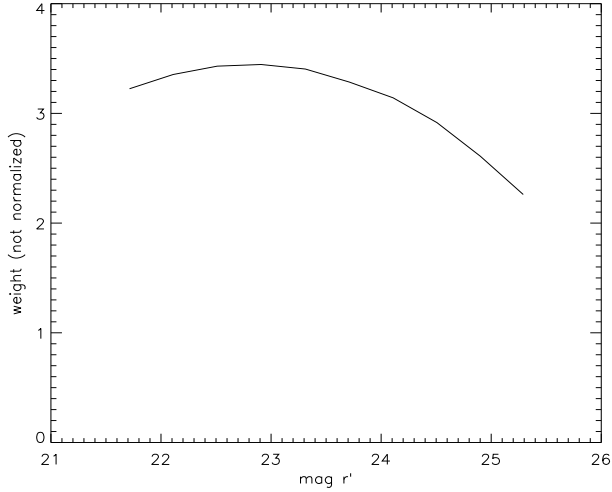}
\caption{\label{magweight} Plot of mean weight per galaxy as
function of magnitude $21.5< r'_{AB} < 25.5$.}
\end{center}
\end{figure}

\begin{figure}[!ht]
\begin{center}
\includegraphics[width=9cm,height=7cm]{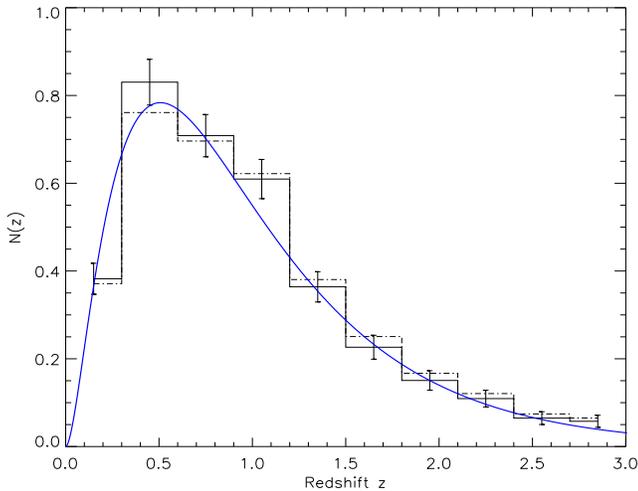}
\caption{\label{HDF} The histogram shows the photometric redshift
distribution of $21.5< r'_{AB} < 25.5$ galaxies of Hubble Deep Field
North and South used in this work. The central solid line is the best
fit model. The solid line histogram is that magnitude weighted
redshift distribution.  The dashed-dot histogram shows the redshift
distribution if the galaxies were not magnitude weighted.}
\end{center}
\end{figure}

We use the source redshift distribution model of Eq.12 and perform a
$\chi^2$ fit, allowing the parameter $z_s$ to vary. We then identify
the $\pm 1\sigma$ and $\pm 2\sigma$ uncertainties, which we
marginalized over in the cosmological parameter estimation. We find
$\alpha= 1.9833$, $\beta=0.6651$,
$z_s=0.0981^{+0.013+0.021}_{-0.011-0.016}$. 
Figure \ref{magweight} shows the unnormalized weight in magnitude
slices in the Deep catalogues. The effect of down-weighting faint
galaxies is taken into account in the source redshift
estimation. Figure \ref{HDF} shows the best fit model and the
underlying photometric redshifts from the Hubble Deep Fields (solid
line). Error bars are Poisson errors. The dashed-dotted line on Figure
\ref{HDF} shows the redshift distribution one would have if we ignore
the weighting. The best fit redshift distribution model has a mean
source redshift of $\approx 1.01$, nearly $0.2$ higher in $z$ than the
Wide survey (Hoekstra et al. 2005).

\begin{figure}[!ht]
\begin{center}
\includegraphics[width=9cm,height=7cm]{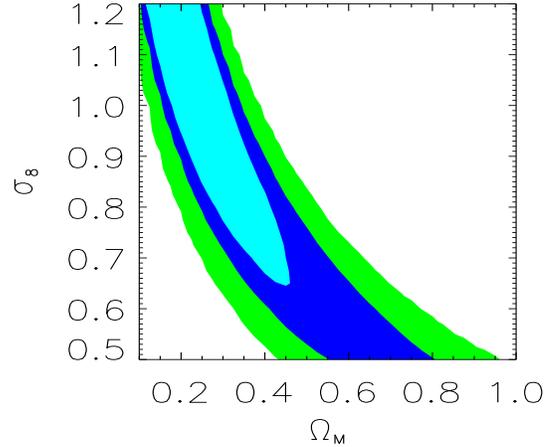}
\caption{\label{omegasigma8_Deep_PandD.ps} $\Omega_m$ and $\sigma_8$
constraints with the Deep data only. The contours show 0.68, 0.95 and 0.999
confidence regions. Errors include
statistical, covariance and residual systematic contributions. The
models are pure Cold Dark Matter fit to the data, marginalized over
the redshift distribution (see Section 7.2 for the details).}
\end{center}
\end{figure}

The constraints on $\Omega_m$ and $\sigma_8$ are obtained after
marginalization of the reduced Hubble constant $h\in[0.6,0.8]$ and
over the $\pm~2\sigma$ limits of the source redshift parameter
$z_s$. The resulting constraints in the $\Omega_m$-$\sigma_8$ plane
are given in Figure \ref{omegasigma8_Deep_PandD.ps}. This figure shows
that the CFHTLS Deep field gives constraints as good as previous
lensing measurements, despite its small field of view.  This is the
consequence of the larger fraction of high redshift galaxies, which
are more strongly lensed. Using the Peacock \& Dodds (1996) non-linear
scheme, we obtain $\sigma_8=0.94\pm 0.15\pm 0.20$
($\pm~1\sigma\pm~2\sigma$) for $\Omega_m=0.3$. Error bars are the one
and two $\sigma$ errors respectively. The Smith et al. (2003) halo
model gives $\sigma_8=0.90\pm 0.14\pm 0.20$, which agrees with
previous normalization measurements. The similarity between
the result obtained using the Peacock \& Dodds and that using the halo
fitted model is not surprising. Indeed, on scales $\gsim 1^{\prime}$,
which dominate our signal, the difference between the two models of
power spectrum is $\lsim 5 \% $. On smaller scales, we would expect an
increasing discrepancy between these different ways to estimate
$\sigma_8$.

\begin{figure}[!ht]
\begin{center}
\includegraphics[width=9cm,height=7cm]{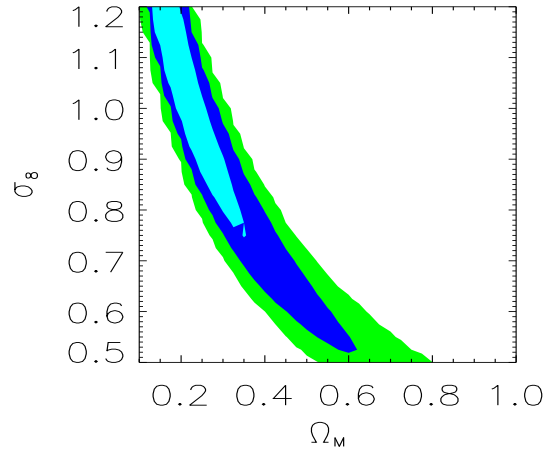}
\caption{\label{omegasigma8_W1+W3+Deep_PandD.ps} Same as Figure
\ref{omegasigma8_Deep_PandD.ps}, combined with the CFHTLS Wide data
(Hoekstra et al. 2005). For $\Omega_m=0.3$ we have
$\sigma_8=0.86\pm0.05$ at $1\sigma$ (see Section 7.2 for details of
the error calculation).}
\end{center}
\end{figure}

We then measure $\sigma_8$ by combining these constraints with those
obtained on the CFHTLS Wide survey (see Hoekstra et al. 2005 for the
details). The result of this joint analysis is shown in Figure
\ref{omegasigma8_W1+W3+Deep_PandD.ps}, and remarkably, the
$\Omega_m$-$\sigma_8$ degeneracy is partially broken. This is the
consequence of measuring the large and small scales simultaneously, as
shown in \cite{JS97}. For $\Omega_m=0.3$, we get $\sigma_8= 0.89\pm
0.06 \pm 0.12$ using Peacock \& Dodds (1996) for the non-linear scheme
and $\sigma_8=0.86 \pm 0.05 \pm 0.11$ for the halo model (Smith et
al. 2003).

The power spectrum normalization is in very good agreement with
results from medium-redshift and low-source-redshift weak lensing
surveys (\cite{HOEK02,VWM05b}). It is remarkable that the parameters
of the redshift distribution, which have been estimated from a
different survey, are such that the normalization $\sigma_8$ lies
within the errors of previous measurements. This is strong evidence
that deep, medium and shallow lensing surveys are in cosmological
agreement, hence reinforcing the ability of cosmic shear to probe the
mass distribution at different redshifts and different scales.

\begin{figure}[!ht]
\begin{center}
\includegraphics[width=9cm,height=7cm]{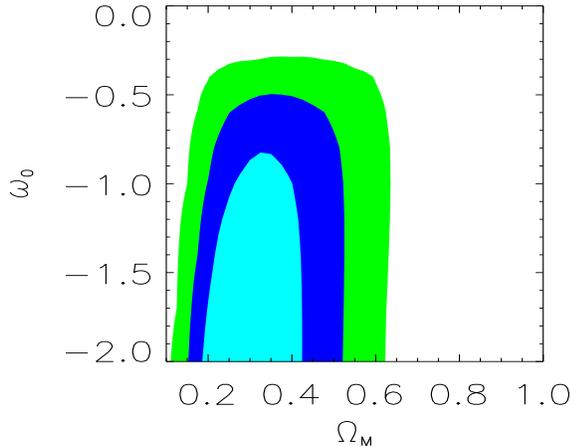}
\caption{\label{DE_Deep_PandD.ps} Dark energy constraints from the
Deep data only. Hidden parameters are marginalized using a flat prior
over $\sigma_8\in[0.7,1.0]$, $h\in [0.6,0.8]$ and within the $\pm
2\sigma$ boundaries of the redshift parameter $z_s$ (see Section
7.2).}
\end{center}
\end{figure}

Weak lensing can also be used to constrain dark energy. Figure
\ref{DE_Deep_PandD.ps} show the upper limit on $w_0$, the constant
equation of state parameter derived from the Deep data only. Here we
used only the Peacock \& Dodds non-linear prescription (a detailed
discussion on non-linear power spectrum correction in the context of
Dark Energy can be found in \cite{HHal05}, which also includes a joint
analysis of the Wide and Deep data). We obtain $w_0 < -0.8$ at
$1\sigma$, and the contours show that this result is independent of
$\Omega_M$. This is particularly interesting because lensing combined
with either cosmic microwave background (Jarvis et al. 2005) or
supernovae will provide a strong constraint on the dark energy
equation of state.

\section{Summary and conclusion}
This paper describes the first cosmic shear studies of CFHTLS Deep
data using the T0001 CFHTLS release.  It uses data collected in $u^*$,
$g'$, $r'$, $i'$ and $z'$ with {\sc Megaprime}/{\sc Megacam} over the
first year of the survey. Only between 1\% and 15\% of the Deep data are therefore in hand depending on the field and on the filter, and so the survey is still 2-3 magnitude below the final goal.

The T0001 data have been used to assess the capabilities of {\sc
Megaprime}/{\sc Megacam} and to clarify the potential and the science
drivers of the CFHTLS Deep survey for weak lensing studies.

The correction for PSF anisotropy works very well, showing that
residual systematics are almost zero at all scales probed by a {\sc
Megacam} field. This is confirmed by the star-galaxy cross-correlation
analysis. This also demonstrates that the CFHT-{\sc Elixir}-{\sc
Terapix} calibration/reduction pipelines can deliver co-added images
which have the required lensing quality.  However, the presence of
B-modes by weak objects at small scales should be further
investigated.

The cosmic shear signal has been detected in the $r'$-band.  Its
consistency and achromaticity has been checked by independent $r'$-
and $i'$- analysis of the same data sets. We have presented results
for three standard two-point shear statistics.

Thanks to the depth of the CFHTLS Deep sample, and using the
photometric redshifts derived from the $u^*$, $g'$, $r'$, $i'$ and
$z'$ images, the galaxy sample was split into low- and high-redshift
sources, and the cosmic shear signal was measured on the two
subsamples separately. Both subsamples show zero B-modes and the shear
amplitude of the high-$z$ sample is clearly higher than the low-$z$
one, with a ratio in agreement with the cosmic shear predictions.  The
amplitude of the signals from the two subsamples are different from
each other at all scales with a significance level higher than 5-sigma
and their shapes follow theoretical expectations of $\Lambda$-CDM
dominated universe.  This strong evidence for the cosmological nature
of the signal shows that the CFHTLS Deep data will allow us to explore
the growth rate of cosmic shear signal with redshift, and hence the
evolution of the dark matter power spectrum as function of lookback
time.

Using only Deep data, and marginalizing over $h$ and the redshift of
sources, we have derived constraints on $\sigma_8$ and $\Omega_m$. We
show that the degeneracy between these two parameters is partially
broken when the analysis is combined with data from Wide
survey. Assuming $\Omega_m=0.3$, we found that $\sigma_8 =0.89 \pm
0.06$ for P\&D and $\sigma_8 =0.86 \pm 0.05$ with the halo model, in
excellent agreement with \cite{VWM05al} ($\sigma_8 =0.83 \pm 0.07$ )
and \cite{HOEK02} ($\sigma_8 =0.86 \pm 0.05$ ).  Likewise, we derive
$w_0<-0.8$ using Deep data alone (see \cite{HHal05} for a deep+wide
analysis).

Our results show that everything is in place to make a full scientific
use the CFHTLS lensing data, and that soon with deeper Deep survey
data and wider Wide survey data, we will able to provide the best
cosmological constraints from weak lensing to date. In
particular, we expect to explore the growth rate of structure from a
tomographic cosmic shear measurement, and to better constrain
cosmological models from the non-Gaussian features derived from a
joint analysis of two-point and three-point statistics. The analysis
of three-point statistics in CFHTLS data goes beyond the scope of this
paper. Indeed, the Deep data used for this paper, is not wide enough
for such a measurement. However, three-point statistics will be
investigated using future samples both for the Deep and Wide survey.

{\acknowledgements We warmly thank the CFHT, {\sc Terapix} and CADC
staff for their assistance and the considerable work they do to
produce the CFHTLS data.  We thank J. Benjamin, F. Bernardeau,
T. Erben, B. Fort, C. Heymans, B. M\'enard, P. Schneider, R. Pell\'o,
C. Schimd, C. Shu, J.-P. Uzan for useful discussions. RM thanks the
City of Paris and IAP for funding his research grants at IAP. YM, ES,
IT and LF thanks the CNRS-INSU and the French Programme National de
Cosmologie for their support to the CFHTLS cosmic shear program. ES
thanks the University of British Columbia for hospitality. LF thanks
the "European Association for Research in Astronomy" training site
(EARA) and the European Community for the Marie Curie doctoral
fellowship MEST-CT-2004-504604. LVW, HH and MJH are supported by the
Natural Sciences and Engineering Research Council (NSERC), the
Canadian Institute for Advanced Research (CIAR) and the Canadian
Foundation for Innovation (CFI).}

\end{document}